\documentclass[preprint,sort&compress]{elsarticle}

\usepackage{lineno,hyperref}
\modulolinenumbers[5]

\journal{Annals of Physics}

\pdfoutput=1
\usepackage{graphicx}
\usepackage{xcolor}
\usepackage{bbm}
\usepackage{amssymb}
\usepackage{amsmath}
\usepackage{tabularx}
\usepackage{ulem}
\usepackage{dsfont}
\usepackage{multirow}
\usepackage{dcolumn}
\usepackage{comment}
\usepackage{float}
\usepackage{footmisc}

\def\0#1#2{\frac{#1}{#2}}

\def\s0#1#2{\mbox{\small{$ \frac{#1}{#2} $}}}

\newcommand{\beq}{\begin{equation}}
\newcommand{\eeq}{\end{equation}}
\newcommand{\bea}{\begin{eqnarray}}
\newcommand{\eea}{\end{eqnarray}}

\newcommand{\f}[2]{\frac{#1}{#2}}

\begin{document}

\begin{frontmatter}

\title{Quantum anomaly and thermodynamics of one-dimensional fermions with antisymmetric two-body interactions}

\author[camblongaddress]{H. E. Camblong}
\author[uhaddress]{A. Chakraborty\corref{mycorrespondingauthor}} \cortext[mycorrespondingauthor]{Corresponding author}
\author[uhaddress]{W. S. Daza}
\author[UNCaddress]{J. E. Drut}
\author[uhaddress]{C. L. Lin}
\author[uhaddress,riceaddress]{C. Ord\'o\~nez}

\address[camblongaddress]{Department of Physics and Astronomy, University of San Francisco, San Francisco, CA 94117-1080 USA}
\address[uhaddress]{Physics Department, University of Houston, Houston, Texas 77024-5005, USA}
\address[UNCaddress]{Department of Physics and Astronomy, University of North Carolina, Chapel Hill, North Carolina 27599-3255, USA}
\address[riceaddress]{Department of Physics and Astronomy, Rice University, MS 61, 6100 Main St., Houston, TX 77005.}

\begin{abstract}
A system of two-species, one-dimensional fermions, with an attractive two-body interaction of the derivative-delta type, features a scale anomaly. In contrast to the well-known two-dimensional case with contact interactions, and its one-dimensional cousin with three-body interactions (studied recently by some of us and others), the present case displays dimensional transmutation featuring a power-law rather than a logarithmic behavior.  We use both the Schr\"{o}dinger equation and quantum field theory to study bound and scattering states, showing consistency between both approaches. We show that the expressions for the reflection $(R)$ and the transmission $(T)$ coefficients of the renormalized, anomalous derivative-delta potential are identical to those of the regular delta potential. The second-order virial coefficient is calculated analytically using the Beth-Uhlenbeck formula, and we make comments about the proper $\epsilon_B\rightarrow 0$ (where $\epsilon_B$ is the bound-state energy) limit. We show the impact of the quantum anomaly (which appears as the binding energy of the two-body problem, or equivalently as Tan's contact) on the equation of state and on other universal relations. Our emphasis throughout is on the conceptual and structural aspects of this problem.
\end{abstract}


\end{frontmatter}

\section{Introduction}

The study of manifestations of scaling SO(2,1) anomalies in nonrelativistic systems has received considerable attention in recent years. {Anomalies} appear when a symmetry is present at the classical level, but is broken by quantum fluctuations; the prime examples in nonrelativistic physics are the two-dimensional (2D) Fermi gas with attractive contact interactions~\cite{jackiw1991mab, PitaevskiRosch, Braaten-Platter,Olshanii, HuLiang,Hofmann,Taylor-Randeria} and dipole-bound anions of polar molecules~\cite{Anomaly-molecular}. On the experimental side, ultracold-atom experiments have shed light on the thermodynamic, collective-mode, and transport properties of that 2D system~\cite{Experiments2D2010Observation, 
	Experiments2D2011Observation, 
	Experiments2D2011RfSpectroscopy, 
	Experiments2D2011DensityDistributionTrapped, 
	Experiments2D2011Crossover2D3D, 
	Experiments2D2012RfSpectraMolecules, 
	Experiments2D2012Crossover2D3D,
	Experiments2D2012Polarons1, 
	Experiments2D2012Polarons2,
	Experiments2D2012Viscosity,
	ContactExperiment2D2012, 
	Vale2Dcriteria,
	Experiments2D2014, 
	Experiments2D2015SpinImbalancedGas, 
	Experiments2D2015PairCondensation,
	Experiments2D2015BKTObservation} (see also~\cite{RanderiaPairingFlatLand,PieriDanceInDisk}).
On the theory side, there have been multiple non-perturbative studies of basic ground-state~\cite{Bertaina,ShiChiesaZhang,Gezerlis} and thermodynamic quantities~\cite{LiuHuDrummond,Enss2D,ParishEtAl,AndersonDrut,BarthHofmann}, and transport~\cite{ChafinSchaefer,EnssUrban,EnssShear}.

One-dimensional (1D) Fermi systems exhibiting the same SO(2,1) scaling anomalous behavior have also been recently studied, both by means of an exact mapping between the bound-state sectors of a 1D, three-body system (3 fermion species, or ``flavors") and a 2D, two-body system (2 fermion species), as well as of a zero-temperature quantum field theory approach to the three-body interaction~\cite{PhysRevLett.120.243002,dazaetalQFT1D3Body}. This mapping was also used to investigate consequences of this anomaly in transport phenomena (bulk viscosity) for those 1D and 2D systems~\cite{JeffOrdonez}. For the 1D system in particular, few-body and finite-temperature thermodynamic aspects were studied numerically~\cite{PhysRevA.99.013615, PhysRevA.102.023313} and high-order virial coefficients were investigated~\cite{PhysRevA.102.063630}.

In this paper, we will add another 1D system to the above list: the derivative-delta potential, $\delta'(x)$, that possesses the same classical scaling SO(2,1) symmetry. The existence of this classical invariance for the 
derivative-delta potential motivated us to investigate the possibility of anomalies in the quantization of this system. There is a vast literature on the quantization of this kind of potential (or similar) from a more mathematical point of view~\cite{Seba,Seba2,Albeverio-Gesztesy-etal1,Albeverio-Gesztesy-etal2,Zhao, Albeverio-Gesztesy_controversy, Griffiths,Kurasov:1993,Christiansen-et-al1,Christiansen-et-al2,Kurasov:1996,Coutinho:1997, WidmerThesis, Al-Hashimi, Gadella-Negro-Nieto, gadellaspectroscopy, zolotaryuk}.
In this work, we follow a physically-motivated approach, using the framework of low-energy effective field theory and renormalization~\cite{Lepage1,Lepage2,Weinberg,polchinski1,polchinski2,bira1,bira2,bira3}, and the ensuing concept of dimensional transmutation~\cite{CW,Huang-QFT:82}. Low-energy theories are valid for energies and momenta smaller than a large cutoff $\Lambda$.  The low-energy Lagrangian is written as an infinite number of terms, with only a handful being the important ones for those low energies, so one can ignore the other terms, with the high-energy information indirectly coded into a few $\Lambda$-dependent coupling constants.  One can formally work with a finite cutoff, and then make contact with low-energy data via a renormalization procedure, which renders physical quantities finite in the infinite $\Lambda$ limit.  In the case of theories with classical scale invariance, the scaling SO(2,1) invariant interactions are ``marginal,'' or renormalizable, and the coupling constants are dimensionless, which is consistent with the fact that at the classical level there are no scales in the theory. However, after renormalization and the introduction of an energy scale (in this case the bound-state energy of the ground state), the original symmetry is broken, and the dimensionless coupling constant is ``transmuted" into the bound-state energy.  This remarkable phenomenon of dimensional transmutation, which was discovered originally in particle physics~\cite{CW}, represents a breakdown of naive dimensional analysis~\cite{Stevenson-QFTDA}, and has been applied in several problems in particle, condensed matter, and nuclear physics~\cite{Weinberg73,Einhorn,Huang-QFT:82,Camblong-DTa,Camblong-DTb}. In the physics of ultracold gases, various aspects of scale symmetry breaking driven by dimensional transmutation have been considered in the early Refs.~\cite{Braaten-Platter,Olshanii, HuLiang,Hofmann,Taylor-Randeria}; moreover, this low-energy, renormalization treatment of anomalies has been used recently with success in the description of thermal and transport properties of one- and two-dimensional quantum gases~\cite{PhysRevLett.120.243002, JeffOrdonez, UHetUNC2D}. As far as the derivative-delta potential is involved, heterostructures at the nanoscopic level may provide an approximate realization of a model with this type of local interaction, as in Refs.~\cite{zolotaryuk,gadellaspectroscopy}; moreover, as in the case of the related two-dimensional delta-function potential~\cite{jackiw1991mab}, physical applications in other systems, including quantum gases, may eventually be developed.  

The paper is organized as follows. In Sec.~\ref{Sec:derivative-delta}, we will use two different frameworks to investigate the bound-state and scattering sectors: the Schr\"{o}dinger equation and a quantum field theory treatment, and we will show their equivalence. This equivalence further supports the robustness of the claims made in this paper. With this foundation, in Sec.~\ref{Sec:Results}, we will exhibit the 
corresponding non-trivial, symmetry-breaking results for the second-order virial coefficient and
Tan's contact~\cite{TanContact-Tan1,TanContact-Tan2,TanContact-Tan3,TanContact-additional1,TanContact-additional2,TanContact-additional3}, following the approach of Refs.~\cite{dilational-anomaly-thermo} and \cite{UHetUNC2D}. We will close in Sec.~\ref{Sec:Conclusions} with a summary and conclusions highlighting the role played by renormalization, dimensional transmutation, and the quantum anomaly in this problem. The appendices include background material for comparison with the regular delta-function interaction potential and an overview of universal relations governed by Tan's contact. 

\section{\label{Sec:derivative-delta}Derivative-delta interaction: Renormalization and quantum anomaly}

\subsection{\label{Sec:Ham}Hamiltonian and Symmetries}
The Hamiltonian we study is $\hat H = \hat H_0 + \hat H_I$, where
\begin{gather}
\hat H_0 = \frac{1}{2} \sum_{s=\uparrow,\downarrow} \int dx \; \frac{\partial \psi_{s}^\dagger}{\partial x}\frac{\partial \psi_s}{\partial x} ~,\label{Eq:T}\\
\hat H_I = g \int dx \, dx' \; \delta'(x-x') \hat n_\uparrow(x)  \hat n_\downarrow(x') .\label{Eq:V}
\end{gather}
Here $n_s(x) = \psi_{s}^\dagger(x) \psi_s(x)$ with $s\in\{\uparrow,\downarrow\}$ are the number densities of the fermion fields. We will choose natural units with $\hbar = k_B = m = 1$.
Since the derivative-delta function
$\delta'(x)$ carries dimensions of inverse length squared, while the 1D density has dimensions of inverse length, the bare coupling $g$ is dimensionless, which reflects the classical scale invariance of the action associated with $\hat H$.
As we show below, however, the coupling runs non-trivially with the cutoff, in such a way that the physical coupling (dimensionally-transmuted scale~\cite{Camblong-DTa,Camblong-DTb}) is the binding energy of the two-body problem.

The scale invariance of the system described by the Hamiltonian of Eq.~(\ref{Eq:V}) is part of a larger symmetry group manifested at the level of the action, and which involves position and time reparametrizations. The most general reparametrizations involve
the generators:	the Hamiltonian $H$ (performing time translations), the dilation operator $D$ (performing rescalings of position ${\bf r} \rightarrow \varrho {\bf r}$ and time  $t \rightarrow \tau t$, such that $\tau = \varrho^2$), and the special conformal operator $K$ (performing translations of reciprocal time $1/t$)~\cite{AFF:76, jackiw-CQM1,jackiw-CQM2,jackiw-CQM3,nr-SO(21)1,nr-SO(21)2,nr-SO(21)3,nr-SO(21)4,nr-SO(21)5,nr-SO(21)6,jackiw1991mab}. These define a noncompact Lie algebra for the group $SO(2,1) \approx SL(2,\mathbf{R})$. The symmetry is enforced as invariance of the classical action---though the Lagrangian and Hamiltonian are not invariant by themselves---for the family of systems often called conformal quantum mechanics (CQM)~\cite{AFF:76}. The ensuing relations can be written in generic $d$-dimensional form (as needed for other conformal systems), thus generalizing the 1D results of the Hamiltonian of Eqs.~(\ref{Eq:T})--(\ref{Eq:V}) (i.e., replacing $x \rightarrow {\bf x}$, $\partial_x \rightarrow \nabla$, etc). Specifically, defining the number density
	\begin{equation}
	n({\bf x}) =   \sum_{s=\uparrow,\downarrow} \psi_{s}^\dagger ({\bf x}) \psi_s ({\bf x})
	\end{equation}
	and momentum density
	\begin{equation}
	{\bf j} ({\bf x}) =  - \frac{i}{2} \sum_{s=\uparrow,\downarrow}  \left( \psi_{s}^\dagger ({\bf x}) \,\boldsymbol{\nabla}
	\psi_s ({\bf x}) -  \boldsymbol{\nabla}  \psi_{s}^\dagger ({\bf x}) \, \psi_s ({\bf x}) \right)
	\; ,
	\end{equation}
	the following two operators~\footnote[1]{Several related definitions have been used: here we highlight those of Refs.~\cite{AFF:76} and~\cite{jackiw-CQM1,jackiw-CQM2,jackiw-CQM3} (including the first-quantized version of CQM): $\hat{D}$ and $K$; and of Ref.~\cite{nrCFT-Nishida-Son} (most common in the recent literature of nonrelativistic CFTs): ${D}$ and $C$. The early nonrelativistic field theory references~\cite{nr-SO(21)1,nr-SO(21)2,nr-SO(21)3,nr-SO(21)4,nr-SO(21)5,nr-SO(21)6} use alternative conventions.}
are generators of this larger symmetry~\cite{AFF:76, jackiw-CQM1,jackiw-CQM2,jackiw-CQM3, nr-SO(21)1,nr-SO(21)2,nr-SO(21)3,nr-SO(21)4,nr-SO(21)5,nr-SO(21)6,jackiw1991mab,nrCFT-Nishida-Son}, 
in addition to the Hamiltonian $H$:
	the dilation operator
	\begin{equation}
	\hat{D} = Ht- \frac{1}{2} D 
	\; , \; \; \text{where}  \; \; \;
	D = \int  d^{d} x \, {\bf x} \cdot   {\bf j} (x)
	\end{equation}
	and the special conformal operator
	\begin{equation}
	K=  t^2H - Dt + C
	\; , \; \; \text{where}  \; \; \;
	C = \int d^{d} x \, \frac{1}{2} \left|{\bf x}\right|^2 n ({\bf x})
	\; .
	\end{equation} 
	If there were no anomaly, the 
	commutators that enforce the classical symmetry would be
	\begin{equation}
	\begin{aligned}
	& [\hat{D},H]_{\rm regular}
	= - i  H 
	\; , \; \; \text{or}  \; \; \;
	[D,H]_{\rm regular}
	= 2 i  H
	\\
	& [\hat{D}, K]_{\rm regular}  = i K
	\; , \; \; \text{or}  \; \; \;
	[D, C]_{\rm regular}  = -2 i C
	\\
	& [H,K]_{\rm regular}  = 2 i  \hat{D}
	\; , \; \; \text{or}  \; \; \;
	[H,C]_{\rm regular}  = -  i  D
	\;  .
	\label{eq:naive_commutators}
	\end{aligned}
	\end{equation}
	However, extra terms arise from the quantization of the theory---without these terms, the theory would be scale invariant. In Sec.~\ref{Sec:TBM_BS}, we indeed show that the required renormalization procedure generates a scale associated with a ground state, which affects the scattering variables and other observables. Such scale symmetry breaking in the form of dimensional transmutation implies a nonzero extra term in the commutator $[D,H]$ that describes the scaling behavior of $H$. From the definitions above, this commutator is the integrated spatial trace of the momentum flux tensor $\Pi_{ij}$ (the negative of the stress tensor)~\cite{nrCFT-Nishida-Son,Bergman}:
	$[D,H] = i \int d^{d}x \,  \Pi_{jj}$ (which involves just $j \equiv x$ in the one-dimensional setup of this paper); thus,
	\begin{equation}
	[D,H]
	= i \left( 2   H + \mathcal{A} \right)
	\label{eq:scale-commutator}
	\; ,
	\end{equation}
	where (cf.~Eq.~(\ref{eq:naive_commutators})) the extra term $[D,H]_{\rm extra} $ 
	yields the anomaly operator 
	\begin{equation}
	\mathcal{A} \equiv
	\frac{1}{i}  [D,H]_{\rm extra} = \int d^{d}x \, \Pi_{jj} - 2 H
	\label{eq:anomaly-operator}
	\; .
	\end{equation}
	This scale anomaly was explicitly derived first in Ref.~\cite{Bergman} for scalar fields with contact interactions,
	in  \cite{1stq-CQM-anomaly1,1stq-CQM-anomaly2,1stq-CQM-anomaly3} for the first-quantized theory; and for Fermi gases with contact interactions in 2D using dimensional regularization~\cite{Hofmann}
	and path integral methods~\cite{Fujikawa-2Danomaly1,Fujikawa-2Danomaly2,Fujikawa-2Danomaly3,Fujikawa-2Danomaly4,Fujikawa-2Danomaly5,Fujikawa-2Danomaly6}.
	Moreover, if a statistical thermal average is taken in Eq.~(\ref{eq:anomaly-operator}), the right-hand side per unit volume becomes the important thermodynamic quantity $dP- 2 E/V$, which generates the equation of state of the corresponding quantum gas by measuring the degree of scale symmetry breaking. This insight, and how it affects thermodynamic relations, including the second virial coefficient as a key signature of the anomaly~\cite{UHetUNC2D}, is discussed in Sec.~\ref{Sec:Results}.
	In that section, we further identify the anomaly or nonzero extra term~(\ref{eq:anomaly-operator}) as Tan's contact~\cite{TanContact-Tan1,TanContact-Tan2,TanContact-Tan3}, using the novel dimensional analysis and scaling relations of Ref.~\cite{dilational-anomaly-thermo}.

\subsection{\label{Sec:TBM_BS}The two-body problem: bound states via momentum-space approach}
In this subsection, we probe the 1D derivative-delta interaction for the existence of a bound state using the two-particle Schr\"odinger equation
\beq
\label{Eq:Sch2BDerivativeDelta}
\left[\frac{-1}{2{\bar m}}\frac{d^2}{dx^2} + g \delta'(x)\right] \psi(x)
=
E \psi(x) \; .
\eeq
Here we will use the units $2\bar m = m = 1$, and the coupling constant $g$ is dimensionless.
This equation has been analyzed before by several authors, with various proposals~\cite{Seba}--\cite{Kurasov:1993}, including the insightful overview of Ref.~\cite{WidmerThesis}.
In this paper, we specifically interpret the potential $V(x) = g \delta'(x) $ as representing the derivative of the delta function, whose properties are handled by straightforward use of a momentum representation via Fourier transforms. We begin by identifying the possible existence of a bound state by allowing a running coupling constant.\footnote[2]{We have also developed a real-space, distributional approach to the system considered here, which, upon connecting with the low-energy framework, validates our results.  This distributional approach does lend itself to a better comparison with the more mathematical literature. These results will be reported elsewhere. \label{real space regularization} } This is mathematically equivalent to the early regularized proposal of Ref.~\cite{Seba} and similar to the treatment in Ref.~\cite{Al-Hashimi}. It should be noted that this is also distinctly different from an alternative definition of the ``delta-prime potential''  within the 4-parameter family of self-adjoint extensions of the generalized point interaction~\cite{Albeverio-Gesztesy-etal1,Albeverio-Gesztesy-etal2} and from the derivative-delta potential with fixed coupling as limit of generalized sequences of regularizing potentials, leading to nontrivial transmission coefficients~\cite{Christiansen-et-al1,Christiansen-et-al2}.

The Fourier transform $\tilde{\psi}(p)$ of Eq.~(\ref{Eq:Sch2BDerivativeDelta}) is obtained through multiplication  by $e^{-ipx}$ and integrating over $x$, leading to 
\beq
\label{Eq:psiPDerivativeDelta}
\tilde{\psi}(p) = { g } \frac{\psi'(0) - i p \psi(0)}{p^2 + \epsilon_B} \; ,
\eeq
where we have specialized to the case of bound states with binding energy $\epsilon_B$, such that $E= -\epsilon_B <0$. The Fourier integrals above involve discontinuous functions at the origin, as required by the matching boundary conditions around the singular point~\cite{Griffiths,Gadella-Negro-Nieto}. Thus, writing $\psi (0)$ and $\psi' (0)$ is an abuse of notation. For a representation of the delta function as a limit of a sequence of even functions,
the values $\psi (0)$ and $\psi' (0)$ are in fact the averages $F(0) \equiv \left[ F(0^{-}) + F(0^{+}) \right]/2$, where $F = \psi, \psi'$, as originally proposed in Refs.~\cite{Griffiths,Kurasov:1996}.

Integrating both sides of Eq.~(\ref{Eq:psiPDerivativeDelta}) with respect to $p$ yields
\beq
\label{Eq:psi0psiprime0}
\psi(0) = \frac{g}{2 \sqrt{\epsilon_B}} \psi'(0) \; .
\eeq
Alternatively, we may multiply by $i p$ and then integrate, which gives us access to the analogue of Eq.~(\ref{Eq:psi0psiprime0}) but where $\psi'(0)$ appears naturally on the left-hand side:
\beq
\label{Eq:psiprime0psi0}
\psi'(0) = g \psi(0) \left( \frac{\Lambda}{\pi}  - \frac{\sqrt{ \epsilon_B}}{2} \right) \; .
\eeq
Here we have imposed an ultraviolet momentum cutoff $\Lambda$ since the resulting integral over $p$ is not
convergent (to be contrasted with the standard delta potential case). 
%
%
The system of equations (\ref{Eq:psi0psiprime0}) and (\ref{Eq:psiprime0psi0}) have a non-trivial solution if
\begin{equation}
\det\begin{bmatrix}
1 & \displaystyle  -\frac{g}{ 2 \sqrt{\epsilon_B}}\\
\displaystyle
g \left( \frac{\Lambda}{\pi} - \frac{\sqrt{\epsilon_B}}{2} \right) & -1
\end{bmatrix} = 0 
\; . \label{determinantdeltaprime}
\end{equation}

From the above, it is easy to see how the coupling $g$ runs at large $\Lambda$:
\beq
\label{Eq:1DEBDeltaPrime}
\frac{\epsilon_B}{\Lambda^2} = \frac{ g^4}{4 \pi^2} \; .
\eeq
In other words, $g$ varies with the square root of the cutoff $\Lambda$ rather than logarithmically. The above power-law behavior is to be contrasted with the other well-known anomalous case, namely that of particles in 2D interacting with a conventional delta-function interaction; there, the analogue result is instead~\cite{jackiw1991mab, Petrov}
\beq
\label{Eq:2DEB}
\frac{\epsilon_B^{(2D)}}{\Lambda^2} = e^{4\pi / g}, \text{ with } g<0 \; .
\eeq
Using the above relation(s), one identifies the binding energy $\epsilon_B$ 
as the physical coupling, and as the emerging scale that breaks the original (classical) scale invariance
for the 2D case with standard delta interaction, or 1D with derivative-delta interaction.

To complete the solution for the bound-state case, we require a proper normalization condition for $\tilde \psi(p)$
by integrating over $|\tilde \psi(p)|^2$, which yields (in the large-$\Lambda$ limit),
\beq
\psi(0) = \left ({ \epsilon_B}\right )^{1/4} \; ,
\eeq
such that
\beq
\psi'(0) = \sqrt{\frac{2\Lambda}{\pi}}\sqrt{ \epsilon_B} \; .
\eeq
Finally, Fourier transforming Eq.~(\ref{Eq:psiPDerivativeDelta}) back to coordinate space, we obtain 
the bound-state wave function
\bea
\label{delta-prime_GSWF}
\psi(x) = \psi(0) e^{- |x| \sqrt{ \epsilon_B}} \left [1+ \sqrt{\frac{\pi}{2 \Lambda}} \left (  \epsilon_B \right )^{1/4} \text{sgn}  (x) \right] \; ,
\label{Eq:psi-x_delta-prime_BS}
\eea
where the signum function $ \text{sgn}(x)$ yields values $\text{sgn} (x) =  x/|x|$ for $x \neq 0$ and is zero for $x=0$. 

Some remarks are in order regarding the nature and interpretation of the solution.
The regularization process involves a running coupling given from Eq.~(\ref{Eq:1DEBDeltaPrime}), so that $g \rightarrow 0$
as $\Lambda \rightarrow \infty$. While it would seem that the second term in Eq.~(\ref{delta-prime_GSWF}) is vanishing
by enforcing this limiting procedure, its presence is nonetheless critical for the wave function to formally satisfy the Schr\"{o}dinger equation.
The discontinuous function $\text{sgn} (x)$ is the source that guarantees the existence of a derivative-delta term in the differential equation.

\subsection{\label{Sec:TBM_SS}The two-body problem: scattering states via momentum-space approach}
For the scattering sector of the derivative-delta interaction, we use $E=k^2$, with $k$ real,
to get the momentum-space Schr\"{o}dinger equation,
\begin{equation}
\bigg( p^2  - E\bigg)\tilde{\psi}(p) = g\big[\psi^\prime(0) - ip\psi(0)\big]
\label{momspaceSchrodingerprime}
\; ,
\end{equation}
via the Fourier-transform of Eq.~(\ref{Eq:Sch2BDerivativeDelta}).
The wave function in momentum space can be derived by inversion of the singular operator on the left-hand side of Eq.~(\ref{momspaceSchrodingerprime}), and takes the form
\begin{equation}
\tilde{\psi}(p) = 2\pi \delta(p-k) + \frac{g}{p^2-k^2-i\epsilon}\bigg[\psi'(0)- ip\psi(0)\bigg] \label{ansatz}
\; ,
\end{equation}
where the first term represents the incident wave and the $i\epsilon$ prescription provides the necessary boundary conditions for outgoing scattered waves. 

Equation~(\ref{ansatz}) can be solved by the following procedure aimed at finding the values of the wave function and its first derivative at the origin, i.e., establishing relations between $\psi(0)$ and $\psi'(0)$. The first relationship can be found by integration of both sides of Eq. (\ref{ansatz}) after multiplying by $dp/2\pi$ and taking the limit $\epsilon \rightarrow 0$, thus yielding
\begin{equation}
{\psi}(0) = 1 + \frac{gi}{2 k}\psi'(0). \label{psi0eq}
\end{equation}
A second relationship is obtained by multiplying both sides of Eq.~(\ref{ansatz}) by $ip \, dp/2\pi$ and integrating,
\begin{equation}
\psi'(0) = ik + g\psi(0) \bigg( \frac{\Lambda}{\pi} + \frac{ik}{2} \bigg) \; . \label{psiprime0eq}
\end{equation}
The values of $\psi(0)$ and $\psi'(0)$ are thus given by solving the system of Eqs.~(\ref{psi0eq}) and (\ref{psiprime0eq}), i.e.,
\begin{align}
\psi(0) &= \frac{\pi  k (1 -g/2)}{\pi  g^2 k/4 - i g^2 \Lambda/2 +\pi  k} \label{psi0value}\\
\psi'(0) &= \frac{i \left(\pi  g k^2/2 - i g k \Lambda  +\pi  k^2\right)}{\pi  g^2 k/4 - i g^2 \Lambda/2 +\pi  k} . \label{psiprime0value}
\end{align}
From Eq (\ref{ansatz}), and using a Fourier transform, the wave function becomes
\begin{align}
\psi(x)	&= e^{ikx} + g\bigg[\frac{\psi'(0) i e^{ik|x|}}{2k} + \frac{\psi(0)e^{ik|x|}}{2} \text{sgn} (x) \bigg] \; .\label{finalpsi}
\end{align} 
Now, this scattering wave function~(\ref{finalpsi}) can be compared against the standard asymptotic form (which, for contact interactions, is valid for all $x \neq 0$),
\begin{equation}
\psi(x) = \begin{cases}
e^{ikx} + R e^{-ikx} \hskip 2em x<0\\
T e^{ikx} \hskip 5.2em x> 0
\end{cases}\label{psixdeltaansatz}
\; .
\end{equation}
Equation~(\ref{psixdeltaansatz}) defines $R$ and $T$ as the reflection and transmission amplitude respectively. 

In the regime $x<0$, Eq.~(\ref{finalpsi}) gives
\begin{equation}
\psi(x) = e^{ikx} + g\bigg[\frac{\psi'(0) i e^{-ikx}}{2k} - \frac{\psi(0)e^{-ikx}}{2} \bigg] \; . 
\label{fornegativex}
\end{equation}
The reflection amplitude $R$ can be found by comparing Eq.~(\ref{fornegativex}) with Eq.~(\ref{psixdeltaansatz}), with the result
\begin{align}
R &=  g\bigg[\frac{\psi'(0)i}{2k} - \frac{\psi(0)}{2} \bigg] \\
&= -\frac{\kappa}{\kappa + i k}
\; .
\label{reflection-coeff_derivative-delta}
\end{align}
The last step involves both the regularization of the scattering using Eqs.~(\ref{psi0value}) and (\ref{psiprime0value}) in the limit $\Lambda \rightarrow \infty$, and its renormalization substituting the bound-state wave number relation $\kappa =\sqrt{\epsilon_B}= \Lambda g^2/2\pi$.

Similarly, in the regime $x>0$,
\begin{equation}
\psi(x) = e^{ikx} + g\bigg[\frac{\psi'(0) i e^{ikx}}{2k} + \frac{\psi(0)e^{ikx}}{2} \bigg] \; .\label{forpositivex}	
\end{equation}
The transmission amplitude $T$ can be found by comparing Eq.~(\ref{forpositivex}) with Eq.~(\ref{psixdeltaansatz}), with the result
\begin{align}
T &= 1 + g\bigg[\frac{\psi'(0)i}{2k} + \frac{\psi(0)}{2} \bigg]\\
&= \frac{ik}{\kappa+ik}
\; .
\label{transmission-coeff_derivative-delta}
\end{align}
Here, we have used the same limits and values defining the regularization and renormalization of the scattering, as for the reflection amplitude. It should be noted that the transmission amplitude and the reflection amplitude are related by the expression 
\begin{equation}
T = 1+R \; ,
\label{Eq:T-R-relation}
\end{equation}
which suggests isotropic scattering. This property can be seen by noting that Eq.~(\ref{psixdeltaansatz}) can be 
recast in a compact form, valid for both regions, by eliminating $T$ with Eq.~(\ref{Eq:T-R-relation}), i.e.,
\begin{equation}\label{EqnSpecific}
\psi(x)=e^{i k x}+R e^{i k|x|} \; .
\end{equation}
On the other hand, the most general (anisotropic) form of the scattering wave function can be written as
\begin{equation}\label{EqnGeneral}
\psi(x)=e^{i k x}+f(\hat{x}) e^{i k|x|},
\end{equation}
where the scattering amplitude $f(\hat{x})$ depends on the one-dimensional version of the solid angle $\hat{x}=x/|x|$, which is either $0$ or $180$ degrees.  Comparing Eqs.~\eqref{EqnSpecific} and \eqref{EqnGeneral},  we conclude the scattering is of the s-wave type.
Indeed,	starting from 
	\begin{equation}
	\begin{aligned}
	\psi(x)&=e^{i k x}+R e^{i k|x|}\\ 
	&=e^{i k x}-\frac{\kappa}{\kappa + i k} e^{i k|x|} 
	\; , 
	\end{aligned}
	\end{equation}	
	one can directly calculate the partial wave decomposition from the general formula
	\begin{equation}	
	\psi_\text{\textbf{k}}(x)=\sum_{\ell=0,1}(-i)^\ell (\text{sgn}(x))^\ell e^{i \delta_\ell }\cos(k |x| - \pi \ell/2+ \delta_\ell)
	\; ,
	\end{equation}	
	which is the 1D version of the general result for the $d$-dimensional partial wave expansion---see Refs.~\cite{eberly1965quantum,Levinson-virial}. This procedure yields	
	\begin{equation}
	\begin{aligned}
	\psi_\text{\textbf{k}}(x) &=
	\f{\displaystyle i \sqrt{k^2+\kappa^2} }{ik+\kappa}
	\cos\left(k|x|+\delta_0\right)
	\\   
	&+i \frac{x}{|x|} \cos\left(k|x|-\f{\pi}{2}+\delta_1\right) \; ,
	\end{aligned}
	\end{equation}
	where 
	\begin{align}
	\delta_0(k) & =\arctan\left(\frac{\sqrt{\epsilon_B}}{k}\right)
	\label{eq:derivative-delta_phase-shift_0}
	\\
	\delta_1 & =0
	\; .
	\label{eq:derivative-delta_phase-shift_1}
	\end{align}

Finally, it is noteworthy that the scattering expressions coincide with the corresponding results for the delta-function potential---in other words, their scattering data, after renormalization of the derivative-delta potential, are identical. 
(See \ref{app:scattering delta} for comparison purposes.)

Incidentally, as a consequence of the derivation of the explicit forms of the scattering phase shifts~(\ref{eq:derivative-delta_phase-shift_0}) and (\ref{eq:derivative-delta_phase-shift_1}),	we would like to highlight an important result of scattering theory of widespread use: Levinson's theorem. This theorem is known to take different forms according to the spatial dimensionality; and indeed, its one-dimensional version includes modifications of the more familiar three-dimensional results. Specifically, the general form of Levinson's theorem in 1D is
	\begin{equation}
	\delta_{\ell}(0)-\delta_{\ell}(\infty) = \pi \left(  N_{B, \ell}  + \nu_{\ell} \right)
	\end{equation}
	for the two channels $\ell =0,1$ (even/odd), where the extra term is
	\begin{equation}
	\nu_{\ell}
	= \begin{cases}
	-\frac{1}{2}  \text{ \hskip 2.9em for $\ell = 0$, non-critical case}\\
	0   \text{ \hskip 2.9em for $\ell = 0$, critical case}\\
	0   \text{ \hskip 2.9em for $\ell = 1$, non-critical case}\\
	\frac{1}{2}   \text{ \hskip 2.9em for $\ell = 1$, critical case}
	\end{cases}
	\end{equation}
and the critical case is defined by the existence of a half-bound state $\epsilon_B = 0$ (with the non-critical case otherwise, i.e.,  with $\epsilon_B>0$ for all bound states). This form of Levinson's theorem has been shown by several independent methods (Sturm-Liouville theory, partial wave analysis, and S-matrix techniques, among others)~\cite{dong2000levinson,sassoli1994levinson,barton1985levinson,eberly1965quantum}; and we have further confirmed it by spectral function methods~\cite{Levinson-virial}, along with closely related results on the Beth-Uhlenbeck formula for the second virial coefficient (which will be discussed in Sec.~\ref{Sec:Results} below). For both the derivative-delta and ordinary delta-function potentials, the only non-zero phase shifts are for the non-critical, even-parity ($\ell =0$) case; thus, $\delta_{0}(0)-\delta_{0}(\infty) = \pi/2$, which is satisfied by the explicit solution~(\ref{eq:derivative-delta_phase-shift_0}). A closely related theorem is Friedel's sum rule, which thus exhibits a similar modification compared to the more familiar three-dimensional result, as shown in the appendix of Ref.~\cite{Lin-2006_Friedel-sum}.

\subsection{\label{sec:NRQFTDeltaPrime}Scattering matrix elements from exact NRQFT}
In this subsection, we confirm the results obtained in the previous two subsections by using the formalism of nonrelativistic quantum field theory. If the interaction potential is given by Eq.~(\ref{Eq:V}) then we can evaluate all the nonzero Feynman diagram contribution and add them up to obtain an exact result for the T-matrix. 
\begin{figure}
	\centering
	\includegraphics[width=0.45\linewidth]{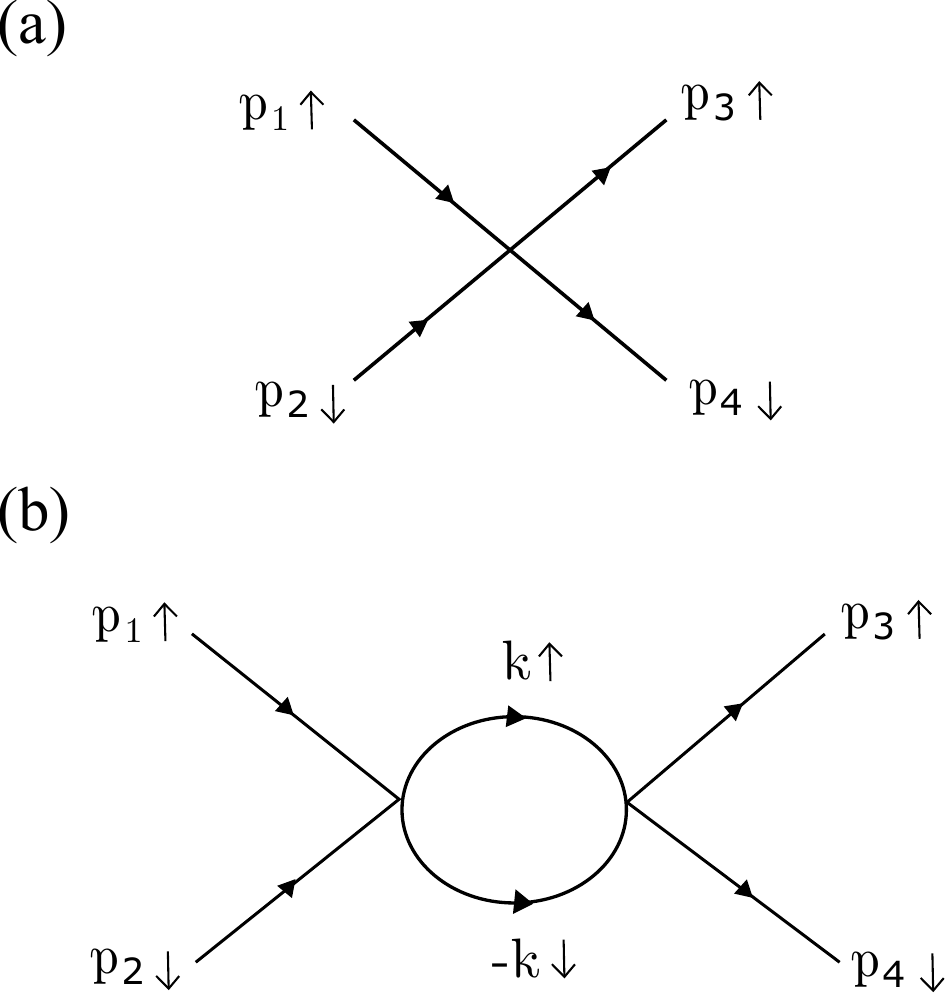}
	\caption{(a) Vertex and (b) 1-loop diagram for the derivative-delta contact potential. Here up and down vertical arrows represent up or down spin and in the COM frame $p_1+p_2 = 0$. So, only $p_1$ and $p_3$ enters the expression of $T_{(1)}$.}
	\label{fig:loop diagrams}
\end{figure}

From Eq.~(\ref{Eq:V}), the tree-level vertex contribution 
$T_{\rm (tree)}(Q,p_1,p_3) $ 
is given by
\begin{equation}
-i \, T_{\rm (tree)}(Q,p_1,p_3) = -ig(ip_1-ip_3)
\; ,
\end{equation}
as shown in Fig.~\ref{fig:loop diagrams}a.
Here, we work in the COM frame of the system and $p_1$, $p_3$ denote the momentum of the spin-up particle. $Q$ denotes the COM energy of the two-body system. To obtain the next-order result, 
$T_{(1)}(Q,p_1,p_3) $, we note that the only nonzero contribution comes from the one-loop diagram (see Fig.~\ref{fig:loop diagrams}b), which gives
\begin{align}
-i  & \, T_{(1)}  (Q,p_1,p_3) 
\nonumber \\
&= \int \frac{dkd\omega}{(2\pi)^2} \, \frac{ig(p_1-k)}{\omega-k^2/2+i\epsilon} \; \frac{ig(k-p_3)}{(Q-\omega)-k^2/2+i\epsilon}  \label{eq:1-loop integral} \\
& =\frac{g^2}{2\sqrt{Q}}\left(\frac{2i\Lambda\sqrt{Q}}{\pi} - p_1p_3 - Q \right) \; . \label{eq:T1}
\end{align}
In Eq.~(\ref{eq:T1}), we have used the expression of the free NRQFT propagator $i(\omega-k^2/2+i\epsilon)^{-1}$ and have introduced a momentum cutoff $\Lambda$ to renormalize the integral in Eq.~(\ref{eq:1-loop integral}). The notation $T_{(n)}$ represents the $T$-matrix contribution of the n-loop diagram.

Using the same procedure, we can now calculate contributions from all higher-order loops recursively. Adding the values of the geometric series of all even-order loops (corresponding to odd-order powers of the coupling $g$), we get
\begin{equation}
\sum_{n=0}^{\infty} T_{(2n)}(Q,p_1,p_3) = 
\frac{i g(p_1-p_3)}{1-\frac{i\Lambda g^2}{2\pi \sqrt{Q} }  \label{eq:even T sum}} \; .
\end{equation}
Similarly, summing up all odd-order loop contributions (corresponding to even-order powers of the coupling $g$),
\begin{equation}
\sum_{n=0}^{\infty} T_{(2n+1)}(Q,p_1,p_3) = \frac{i g^2}{2\pi\sqrt{Q}}\frac{\left(2i\Lambda\sqrt{Q} - \pi p_1p_3 - \pi Q \right)}{1-\frac{i\Lambda g^2}{2\pi \sqrt{Q}}} \; .
\label{eq:odd T sum}
\end{equation}
In all the denominators, $\Lambda$ should actually read $\Lambda +i \pi \sqrt{Q}/2$, but the extra term has been dropped for convenience in the limit when it is taken to be large.
To renormalize these expressions, we now note that Eqs.~(\ref{eq:even T sum}) and (\ref{eq:odd T sum}) have a unique pole at $Q=-\epsilon_B$, whence the bound-state energy $\epsilon_B$ is given by the equation
\begin{equation}
1-\frac{\Lambda g^2}{2\pi \sqrt{\epsilon_B}}=0 \implies
\epsilon_B = \frac{\Lambda^2 g^4}{4\pi^2}  \label{eq:1DEBDeltaPrimeQFT}
\; .
\end{equation}
Equation~(\ref{eq:1DEBDeltaPrimeQFT}) matches exactly with Eq.~(\ref{Eq:1DEBDeltaPrime}), which was obtained from the quantum-mechanical treatment of the two-body problem. Using the expression of the bound-state energy and taking the cutoff to infinity, we see that the sum of the even-order loop terms vanishes whereas the sum of the odd-order loop terms remains finite and nonzero. 

The exact result for the $T$-matrix is thus completely determined by the sum of the odd-order loop terms, and is given by
\begin{equation}
T_{\rm (exact)} (Q,p_1,p_3)  = - \frac{2\sqrt{\epsilon_B}}{1-\frac{i\sqrt{\epsilon_B}}{\sqrt{Q}}} \label{eq:Texact}
\; .
\end{equation}
It should be noted that, as usual, the scattering amplitude is proportional to the momentum-space elements of the T-matrix, Eq.~(\ref{eq:Texact}). Going back to Eqs.~(\ref{EqnSpecific}) and (\ref{EqnGeneral}), and using the form of the propagator in one dimension, it follows that the reflection coefficient is given by $R= T_{\rm (exact)}(Q,p_1,p_3)/2i \sqrt{Q}$, where $p_1=p_3=\sqrt{Q}$ for on-shell conserved momenta. Thus, with the identifications $k \equiv \sqrt{Q} $ and $\sqrt{\epsilon_{B}} = \kappa$ in Eqs.~(\ref{EqnSpecific}) and (\ref{EqnGeneral}) (and throughout Subsec.~\ref{Sec:TBM_SS}), this implies that Eq.~(\ref{eq:Texact}) also agrees with the basic scattering results and, specifically, with the amplitude coefficients of Eqs.~(\ref{reflection-coeff_derivative-delta}) and (\ref{transmission-coeff_derivative-delta}).

This concludes the discussion of the bound-state and scattering sectors using the NRQFT treatment and corroborates the results found in Subsecs.~\ref{Sec:TBM_BS} and~\ref{Sec:TBM_SS}. 
It should be noted that a similar calculation for the delta-function potential (as shown in \ref{app:1DDeltaQFT}) yields the same expression for the exact $T$-matrix, with identical residue and pole at the bound-state energy given by Eq.~(\ref{eq:1DEBDelta}). 
These findings support the conclusion drawn at the end of the previous subsection that the scattering data of the derivative-delta interaction coincide exactly with that of the delta-function potential. At first sight, this outcome is surprising and calls for a more detailed explanation, as discussed in the next paragraph.
	
The important condition that our findings are forced to satisfy is that	{\it the observable outcomes cannot be extrapolated to arbitrarily high energies.\/} Operationally, the fact that the model we are using is not the delta-function potential---even though the observables appear to be the same---can be interpreted within the effective-field-theory paradigm: the renormalization techniques generate the relevant low-energy physics from whichever model is used. An appropriate analogy is that, if the potential is interpreted as a multipole expansion, then the leading order is the delta-function interaction, followed next by the first-derivative of the delta function. Thus, for an antisymmetric potential, the leading order becomes the derivative-delta potential analyzed in this paper; the ensuing scattering expressions are obtained via renormalization as discussed in this paper, and for sufficiently low energies they generate the same observables as the ordinary delta-function interaction. Moreover, the symmetry analysis shows that these interactions are treated differently: unlike the standard delta interaction, the derivative-delta potential is classically scale invariant. Thus, the dimensional scale governing the final results for both the bound-state and scattering sectors is an instance of dimensional transmutation, which is generated by the need for a running coupling and for renormalization. Additional consequences of this scale symmetry breaking are discussed in the next section.


\section{Results for Interacting Gases: Thermodynamics and Tan's contact\label{Sec:Results}}

\subsection{Second-order virial coefficient} 
\label{sec:virial}

	Thermodynamics plays a central role in the study of interacting quantum gases~\cite{Qgases-review}. Thus, in this section we analyze some of the thermodynamic properties of a Fermi gas with derivative-delta interactions, governed by the Hamiltonian of Eqs.~(\ref{Eq:T})--(\ref{Eq:V}). 
	We have shown in Sec.~\ref{Sec:derivative-delta} that, despite its classical scale invariance,
	this system exhibits symmetry breaking as a quantum anomaly. Moreover, the noninteracting behavior corresponds to a scale-invariant system, as shown in Subsec.~\ref{sec:anomaly-eq-state}.
	Therefore, one may suspect the existence of a key signature of the anomaly as scale symmetry breaking
	in the degree of departure from a scale-invariant behavior. One particularly simple such deviation is afforded by the virial expansions, as discussed next. 
	
	As a reminder, the generically called virial expansions provide a well-known general method to deal with interacting gases, in which the deviations from the scale-invariant, noninteracting, ideal gas law are considered. In their most direct form, they involve expansions in powers of the fugacity $z= e^{\mu \beta}$, where $\beta$ is the inverse temperature and $\mu$ is the chemical potential. In terms of the grand-canonical partition function  $\mathcal{Z}= \sum_{n \geq 0} Q_{n} z^n $, with $Q_n$ being the $n$-particle canonical partition functions, the virial expansion takes the form 
	\begin{equation}
	-\beta  \Omega = \ln \mathcal{Z} = Q_{1} \sum_{n \geq 1} b_n z^n
	\; ,
	\label{eq:virial-expansion}
	\end{equation}
	where $\Omega = - PV$ is the grand potential, $P$ is the pressure and $V \equiv L $ is the volume (here being a one-dimensional length). This defines the virial coefficients $b_n$, which depend on the $n'$-body problems with $2\leq n' \leq n$. In general, this is a non-perturbative expansion applicable in the dilute limit.
	
	The connection between the scale anomaly and the virial expansion outlined above is heuristic. A more detailed analysis can be established by writing the anomaly defined from first principles in Eqs.~(\ref{eq:scale-commutator})--(\ref{eq:anomaly-operator}), whose thermal average yields 
	the thermodynamic quantity $dP- 2 E/V$. This can be directly related to the grand potential $\Omega$, and, therefore, to the virial coefficients, as shown in Subsec.~\ref{sec:Tan-contact}.

Thus, using the analysis of the two-body problem from Sec.~\ref{Sec:derivative-delta}, in combination
with the celebrated Beth-Uhlenbeck formula~\cite{Huang-StatPhys}, we obtain an exact expression for the second-order virial coefficient $b_2$, which we have not been able to find in the extant literature. The explicit connection between the virial coefficients, the Beth-Uhlenbeck formula, Tan's contact and scale anomaly was first established in  Ref.~\cite{UHetUNC2D}. The Beth-Uhlenbeck formula relates the binding energy $\epsilon_B$ and the derivative of the scattering phase shifts $\delta_\ell$ with respect to momentum with the change $\Delta b_2 = b_2 - b_2^{({\rm free})}$ in the second virial coefficient due to interactions (where $b_2^{({\rm free})}$ corresponds to the nointeracting or free system). Specifically,
\beq \label{EqnDeltab2}
\sqrt{2} \Delta b_2 = \sum_j e^{\beta \epsilon_B(j)} + \frac{1}{\pi} \sum_{\ell = 0,1}\int dk \, \frac{d \delta_\ell}{d k} e^{- \beta k^2} \; ,
\eeq
where $\beta$ is the inverse temperature, the sum over $j$ covers all the bound states (there is only one in our case), and the sum over $\ell = 0,1$ 
goes over the even (symmetric, s-wave) and odd (anti-symmetric, p-wave) scattering channels in 1D, respectively. 

For the derivative-delta interaction, we derived the phase shifts  $\delta_0$  and $\delta_1=0$ in Eqs.~(\ref{eq:derivative-delta_phase-shift_0}) and (\ref{eq:derivative-delta_phase-shift_1}).
Then, Eq.~(\ref{EqnDeltab2}) gives
\begin{equation}
\sqrt{2} \Delta b_2=\frac{1}{2}e^{\beta \epsilon_B}\left(1+\text{erf}\left(\sqrt{\beta \epsilon_B}\right) \right) \; .
\label{eq:virialdeltaprime}
\end{equation}
Notice that $\lim \limits_{\epsilon_B \rightarrow 0} \sqrt{2}\Delta b_2=1/2$, and not zero as expected (free case limit). Applying the Beth-Uhlenbeck formula naively for the delta-function potential yields the same incorrect limit (since $\delta_0(k)$ is the same). This issue can be resolved by carefully treating the subtleties of the 1D case~\cite{Amaya-Tapia,Levinson-virial}. It has been shown using spectral-density or path-integral methods that the modified formula for 1D for a delta-function potential is 
\begin{equation}
\sqrt{2} \Delta b_2=-\frac{1}{2}+\frac{1}{2}e^{\beta \epsilon_B}\left(1+\text{erf}\left(\sqrt{\beta \epsilon_B}\right) \right) \label{eq:virialdelta} \; ,
\end{equation}
which gives $\Delta b_2=0$ in the $\epsilon_B\rightarrow 0$ limit. The origin of the extra $-1/2$ term can be attributed to the infrared divergence of the density of states and hence of the scattering part of $\Delta b_2$. Once this divergence is regularized, one gets the extra term as a contribution of the zero-energy scattering state. The emergence of the $-1/2$ term in 1D is also closely related to the Levinson's theorem as explained in Ref.~\cite{Levinson-virial}, 
and summarized at the end of Subsec.~\ref{Sec:TBM_SS}. Since the scattering data of the derivative-delta potential are the same as the delta potential, we expect the same relation (\ref{eq:virialdelta}) to hold true for both interactions.


\subsection {Beyond the second order in the virial expansion}

As discussed in the previous subsection, the virial expansion at order $n$ involves the physics of the 2- through $n$-body problems that drive the many-body physics. In particular, as we have just computed, the Beth-Uhlenbeck formula encodes the 2-body physics. To go beyond the second order in the virial expansion, one has to use the expansion of the logarithm $\ln {\mathcal Z}$ to derive the virial coefficients $b_n$ in terms of the $n$-particle canonical partition function $Q_{n}$. The familiar formulas start with $n=3$, given by
\beq
\Delta b_3 = \frac{\Delta Q_3}{Q_1} - Q_1 \Delta b_2 \; ,
\eeq
where the change in $Q_3$ due to interactions is given by
\beq
\Delta Q_3 = 2\Delta Q_{21} \; ,
\eeq
with $Q_{21}$ being the canonical partition function for 2 particles of one species and 1 particle of the other.

Using the second-quantized form of the kinetic and potential energy operators, Eqs.~(\ref{Eq:T}) and (\ref{Eq:V}), 
one may evaluate the above by carrying out a leading-order semiclassical approximation along the lines of Ref.~\cite{ShillDrut},
such that $e^{-\beta \hat H} \simeq e^{-\beta \hat T} e^{-\beta \hat V}$. 
Within such an approximation, and using a complete set of two-particle states to evaluate $Q_{21}$, we obtain
\beq
\Delta b_3 = -\sqrt{2} \Delta b_2 \; .
\label{eq:delta-b3}
\eeq
%
The remarkably simple result of Eq.~(\ref{eq:delta-b3}) shows that the change in the third virial coefficient can be reduced to the change in $b_2$, thus also being governed by the anomalous breaking of scale symmetry.

\subsection{Anomaly in the equation of state} 
\label{sec:anomaly-eq-state}

A direct measure of the anomaly for a quantum gas is provided by an appropriate statistical thermal average of Eq.~(\ref{eq:anomaly-operator}).
Clearly, this measures the degree of scale symmetry breaking and identifies the quantity 
\begin{equation}
\frac{{\mathcal A} }{ V} = dP- 2 \frac{E}{V}
\label{eq:anomaly-thermo}
\end{equation}
as the anomaly density (in a $d$-dimensional space). In this subsection and the next one, we evaluate the anomaly arising from the derivative-delta interaction by the general nonperturbative method developed in Ref.~\cite{dilational-anomaly-thermo}.

In truly scale invariant systems, such as noninteracting ones, the pressure $P$ may be written in terms
of the inverse temperature $\beta$ and the chemical potential $\mu$ as $P = \beta^{\alpha} f(\beta \mu)$,
where $\alpha = -d/2 - 1$.
Here, $f$ stands for a generic dimensionless function with a dimensionless argument that encodes the functional dependence of the pressure, once the temperature scaling is factored out. The advantage of isolating the dependence on the dimensionful parameter $\beta$ is that one readily derives, using thermodynamic identities and partial differentiation with respect to $\beta$ and $\mu$, the well-known result 
\beq
\label{Eq:ScaleInvEOS}
P = \frac{2}{d}\frac{E}{V} \; ,
\eeq
where $E$ is the total energy and $V$ is the $d$-dimensional volume. In scale-anomalous systems like the one put forward here, the pressure acquires a second physical, dimensionless 
parameter via the anomaly, which we will write as $\beta \epsilon_B$. Therefore, dimensional analysis implies the existence of a dimensionless function $f$ such that 
\begin{equation}
P = \beta^{\alpha} f(\beta \mu, \beta \epsilon_B)
\; ,
\label{eq:pressure-dependence}
\end{equation} 
where the function $f$ has an additional dimensionless argument. 
(By abuse of notation, we use the same symbol $f$ as in the absence of an anomaly.) Following the derivation outlined above, 
\beq
\label{Eq:ScaleAnomEOS}
P - \frac{2}{d}\frac{E}{V} =  \frac{2}{d}\beta^{\alpha} \frac{\partial f}{\partial (\beta \epsilon_B)} \beta \epsilon_B = 
\frac{2}{d}\beta^{\alpha} \frac{\partial f}{\partial \ln (\beta \epsilon_B)} 
\; ,
\eeq
which shows that the emergence of the dimensionful parameter results in a contribution to the equation of state that breaks the scale invariant result of Eq.~(\ref{Eq:ScaleInvEOS}). 
Equation~(\ref{Eq:ScaleAnomEOS}) is clearly proportional to the anomaly density ${\mathcal A}/V$; a more detailed analysis follows in the next subsection. Moreover,
the existence of this second dimensionless parameter $\beta\epsilon_B$, in addition to  $\beta \mu$, is again a manifestation of dimensional transmutation~\cite{CW,Huang-QFT:82,Stevenson-QFTDA}: the emergence of a finite dimensionful parameter or scale $\epsilon_B$ in the quantization process (regularization and renormalization) having started with a classical theory that only had a dimensionless coupling constant $g$. In addition, the emergent dimensionful parameter $\epsilon_B$ can be combined with $\beta$ to form the new dimensionless quantity $\beta\epsilon_B$. Thus, in practice, for the many-body system, the original coupling $g$ disappears after regularization and renormalization, and is transmuted into a new dimensionless parameter $\beta\epsilon_B$. This can be interpreted as a kind of coupling constant for the many-body system~\cite{UHetUNC2D}. In the case of few-body physics, the new parameter is often of the form $k/\sqrt{\epsilon_B}$ or $\sqrt{\epsilon_B} x$, 
as shown in Sec.~\ref{Sec:derivative-delta}. These conclusions involve the generalized form of dimensional analysis discussed in Refs.~\cite{Stevenson-QFTDA} and~\cite{Camblong-DTa,Camblong-DTb}, and provide additional support for the nontrivial physics consequences implied by the derivative-delta interaction in one dimension.

\subsection{Anomaly as Tan's contact}
\label{sec:Tan-contact}

Tan's contact, originally derived in~\cite{TanContact-Tan1,TanContact-Tan2,TanContact-Tan3}, governs various properties of many-body quantum systems, giving a set of exact universal relations~\cite{TanContact-additional1,TanContact-additional2,TanContact-additional3,Valiente1,Valiente2,WernerCastin,ContactReview1,ContactReview2}. Insightful derivations of Tan's contact from quantum field theory were introduced in Ref.~\cite{Braaten-Platter} and Ref.~\cite{Hofmann}. In our derivation below, we follow the approach of Refs.~\cite{dilational-anomaly-thermo} and \cite{UHetUNC2D}. All of these methods display the feature that a finite quantity governing universal relations, i.e., Tan's contact, can be built from the bare quantities in the interaction Hamiltonian.

In our case, $d=1$ and the anomalous term in Eq.~(\ref{Eq:ScaleAnomEOS}) is indeed Tan's contact. 
Since $\beta P V = \ln \mathcal Z$, where $V=L$ is the volume and 
$\mathcal Z = \text{Tr} \exp\left[- \beta (\hat H - \mu \hat N)\right]$ is the grand-canonical partition function, 
the only way in which $f$ can depend on $\epsilon_B$ is through the dimensionless bare coupling $g$ that appears in $\hat H$:
\beq
\frac{\partial f}{\partial \ln(\beta \epsilon_B)} =
\frac{\sqrt{\beta}}{L} \frac{\partial \ln \mathcal Z}{\partial g} \frac{\partial g}{\partial \ln(\beta \epsilon_B)} 
\label{Eq:ScaleAnomEOS-aux}
\eeq
where
\begin{align}
\frac{1}{\beta L}\frac{\partial \ln \mathcal Z}{\partial g} &= 
\frac{1}{L}\int dx \left\langle \frac{d\hat n_\uparrow(x)}{dx}  \hat n_\downarrow(x)\right\rangle \nonumber \\&= 
\left\langle \frac{d\hat n_\uparrow(0)}{dx} \hat n_\downarrow(0) \right\rangle \; ,
\end{align}
and the angle brackets denote a thermal expectation value in the grand-canonical ensemble; we also assume the system is spatially homogeneous. Thus, for our scale-anomalous 1D system, the anomaly density~(\ref{eq:anomaly-thermo}) is 
\beq
\label{Eq:ScaleAnomEOS2}
\frac{ {\mathcal A} }{L} =
P - 2\frac{E}{L} =  {\mathcal C} \; ,
\eeq
where we identify Tan's contact density
\beq
\mathcal C \equiv 2 \frac{\partial P}{\partial \ln(\beta \epsilon_B)} =
2\frac{\partial g}{\partial \ln(\beta \epsilon_B)} \left\langle \frac{d\hat n_\uparrow(0)}{dx} \hat n_\downarrow(0) \right\rangle \; .
\label{eq:tanscontact_derivation}
\eeq
for the derivative-delta interaction. Several remarks are in order regarding Eqs.~(\ref{Eq:ScaleAnomEOS2})--(\ref{eq:tanscontact_derivation}). First, Eq.~(\ref{Eq:ScaleAnomEOS2}) has the form anticipated in the discussion following Eq.~(\ref{eq:anomaly-operator})---this is the anomaly as viewed from the extra, symmetry-breaking term in the commutator $[D,H]$. Second, the dimensions of the contact density are those of pressure or energy density, which in 1D amounts to $1/L^3$. Third, as in other cases, the contact has a remarkable structural form that factorizes into a two-body piece: {${\partial g}/{\partial \ln(\beta \epsilon_B)}$, which gives the $\beta$-independent contribution}; and a many-body piece: the thermal expectation value of the double-occupancy operator. For the former, Eq.~(\ref{Eq:1DEBDeltaPrime}) gives 
\beq\label{Eq:contContact}
\frac{\partial g}{\partial \ln (\beta \epsilon_B)} = \frac{g}{4} 
\eeq
in the continuum. It is noteworthy that Eq.~(\ref{Eq:contContact}) is an exact result that includes all orders of perturbation theory, as derived	via an infinite summation of diagrams for the T-matrix in Sec.~\ref{sec:NRQFTDeltaPrime}; and it is the complete nonperturbative result as derived in Sec.~\ref{Sec:TBM_BS}.
Plugging Eq.~(\ref{Eq:contContact}) into Eq.~(\ref{Eq:ScaleAnomEOS2}) gives a contact-term expression similar to the 2D one:
\beq
P - 2\frac{E}{L} = 
\frac{g}{2} \left\langle \frac{d\hat n_\uparrow(0)}{dx} \hat n_\downarrow(0) \right\rangle 
\; .
\label{eq:tanscontactdeltaprime}
\eeq 
The right-hand side of Eq.~(\ref{eq:tanscontactdeltaprime}) involves the product of seemingly individual factors: the bare coupling constant $g$ and the expectation value $\displaystyle \left\langle \frac{d\hat n_\uparrow(0)}{dx} \hat n_\downarrow(0) \right\rangle $; these have singular limits (zero and infinity), but their product is a well-defined finite and nonzero quantity. As discussed at the beginning of this subsection, the identification of a finite, observable quantity from the bare quantities in the interaction Hamiltonian is a standard procedure in various treatments of Tan's contact. Here, the procedure is adapted to the derivative-delta interaction.

Equation~(\ref{eq:tanscontactdeltaprime}) can also be derived, in principle, following the procedure of Refs.~\cite{Hofmann,dazaetalQFT1D3Body}, which was based on the use of dimensional regularization (DR) in a quantum field theory formulation. This method explicitly displays the anomalous algebra of Eqs.~(\ref{eq:scale-commutator})--(\ref{eq:anomaly-operator}), involving the dilation operator $D$. However, for the derivative delta interaction, the use of DR is very subtle---work on this is currently in progress.

The analysis above also shows, inter alia, that there is a direct relationship between the anomaly and the virial expansion, as anticipated in Subsec.~\ref{sec:virial}. 
In effect, from 
the virial expansion~(\ref{eq:virial-expansion}),
with $\Omega = - PV$,
and using Eqs.~(\ref{eq:pressure-dependence}) and (\ref{Eq:ScaleAnomEOS}),
it follows that the anomaly density~(\ref{eq:anomaly-thermo}) in 1D is
	\begin{equation}
	\frac{ {\mathcal A} }{L} = - 2 \epsilon_B \frac{\partial}{\partial \epsilon_B }
	\left( \frac{ \Delta \Omega }{ L} \right)
	= 2 \, \frac{ Q_{1}}{L}  \, \sum_{n \geq 2} z^n  \frac{\partial}{\partial \beta } \Delta b_n
	\; ,
	\label{eq:anomaly-virial-relation}
	\end{equation}
where $\Delta b_n = b_n - b_n^{({\rm free})}$ is the change in $b_n$ due to interactions (as in Subsec.~\ref{sec:virial}). This approach can be further developed, and Eq.~(\ref{eq:anomaly-virial-relation}) is one of the several such relationships that can be established to compute the changes in all virial coefficients~\cite{UHetUNC2D}.
	
As an addendum, given that theories with short-range interactions have a short-distance behavior governed by Tan's contact, a set of universal relations is expected for the derivative-delta system analyzed in this paper. In essence, given the binding energy $\epsilon_B$, the characteristic inverse-length scale is $\sqrt{\epsilon_B}$, and this determines the short-distance asymptotic behavior of the density-density correlation or dimer distribution function $g_{12}^{(2)}$ for each pair of particles of different species. Moreover, the spatially integrated distribution function is proportional to the contact. The implementation and justification of this program is briefly discussed in \ref{sec:Universal relations}. A more complete derivation of universal relations will be given elsewhere.


\section{Summary and Conclusions \label{Sec:Conclusions}}

In this work, we have analyzed the scale anomaly in a quantum system of nonrelativistic particles
with a pairwise derivative-delta interaction. 

After summarizing the SO(2,1) symmetry and anomaly properties of this system, we have fully derived the bound-state and scattering sector properties of generic derivative-delta systems, and their application to the two-body problem, using both Schr\"{o}dinger and NRQT methods. We found that the relevant observables at low energies are the same as for the ordinary delta-function potential. With that information at hand, we derived some key thermodynamic properties of the quantum gas related to the scale anomaly and displayed the emergence of the equation of state via Tan's contact. These results are in agreement with the structural form of the anomaly expected from the SO(2,1) algebra. In addition, we calculated a remarkable signature of the anomaly as the second-order virial coefficient---and, in a semiclassical approximation, the third-order coefficient as well. Finally, we outlined a program for universal relations such as the density-density correlation function.

Moreover, despite their coincident low-energy observables, it is noteworthy that the model interaction that we have used in this paper, i.e., the derivative-delta type, is not strictly identical to the regular delta interaction. This is because it follows the usual renormalization approach with a running coupling and a limiting process. A distinction needs to be made between the bare theory and the renormalized one. The fact that there is such a difference is captured by Eq.~(\ref{Eq:psi-x_delta-prime_BS}) for the bound-state sector and Eqs.~(\ref{fornegativex}) and (\ref{forpositivex}) [to be combined with Eqs.~(\ref{psi0value}) and (\ref{psiprime0value})] for the scattering sector; and likewise, between Eqs.~(\ref{eq:even T sum}) and (\ref{eq:odd T sum}) and their counterparts after renormalization. Most importantly, {\it in the effective-field-theory paradigm, the outcome of this renormalization analysis captures the relevant low-energy physics.\/} This has been interpreted as the analogue of a multipole expansion, where, for an antisymmetric potential, the leading order becomes the derivative-delta potential analyzed in this paper. The role of the effective-field-theory approach is more evident from our derivation of the T-matrix using the two-body non-relativistic field theory diagrams in momentum space. This approach further displays the critical role played by a running coupling in the symmetry breaking process of the SO(2,1) anomaly.

In conclusion, even though the derivative-delta systems do not have a characteristic scale, the renormalization process transmutes the original dimensionless coupling into a relevant physical parameter. This dimensional transmutation produces a theory where the conformal SO(2,1) symmetry is broken by quantum fluctuations: a quantum scale anomaly. We have explicitly shown various manifestations of this anomaly at the level of the thermodynamics,
including both the equation of state and the second virial coefficient. In addition, this system should exhibit the familiar breathing modes~\cite{PitaevskiRosch} in the presence of a harmonic trapping potential, as these rely on the same SO(2,1) symmetry algebra. Moreover, the existence of the scale anomaly would imply a shift in their monopole excitation frequency as predicted in Ref.~\cite{Olshanii}; in principle, this would require further study of Eq.~(\ref{eq:tanscontact_derivation}). Finally, the robustness of this nontrivial physics is further supported by the use of renormalization techniques in coordinate space, as will be discussed elsewhere, including additional consequences of the associated symmetry breaking.

\section*{Acknowledgments}
	\noindent
	We would like to acknowledge the participation of Jonathan Ball in the early phases of this work. This work was supported in part by the U.S. National Science
	Foundation under Grant No. PHY1452635 (Computational Physics Program), 
	the US Army Research Office Grant No. W911NF-15-1-0445,
	and the University of San Francisco Faculty Development Fund. This material is based upon work supported by the Air Force Office of Scientific Research under award FA9550-21-1-0017.


\begin{appendix}
	\section{\label{sec:bound state delta}Bound states for the standard delta interaction}
	In this appendix, we review the standard bound-state solution to the 1D two-particle Schr\"odinger equation for the simple delta-type interaction, namely
	\beq
	\label{Eq:Sch2Bdelta}
	\left[\frac{-1}{2{\bar m}}\frac{d^2}{dx^2} + \lambda \delta(x)\right] \psi(x)
	= E \psi(x).
	\eeq
	We are going to use units $ 2 \bar m = m = 1$, and $g$ carries dimensions of inverse length. Transforming to momentum space via a Fourier transform one obtains
	\beq
	\left ( p^2 - E\right)\tilde{\psi}(p) = -\lambda {\psi}(0) \; .
	\label{Eq:Sch2Bdelta-momentum}
	\eeq
	Solving for $\tilde{\psi}(p)$ and integrating over $p$ one obtains the eigenvalue condition
	\beq
	\label{Eq:EigenvalueCondition}
	\frac{-\lambda}{2\pi} \int \frac{dp}{ p^2 - E} = 1 \; .
	\eeq
	To find the bound states and search for a solution where $E = -\epsilon_B < 0$
	for $\lambda < 0$ (we discuss the scattering states further below). 
	Carrying out the (convergent) integral, we obtain a single bound state
	\beq
	\epsilon_B = \frac{\lambda^2}{4} \; . \label{eq:1DEBDelta}
	\eeq
	If instead of integrating $\tilde{\psi}(p)$, we integrate $|\tilde{\psi}(p)|^2$, then the (bound-state) normalization condition yields
	$\psi(0)$ in terms of $\epsilon_B$ up to an irrelevant phase, namely,
	\beq
	\psi(0) = \sqrt{ \frac{|\lambda|}{2} } = \left ({ \epsilon_B}\right )^{1/4} \; .
	\label{Eq:psi-x-0}
	\eeq 
	This completes the solution of the bound-state wave function in momentum space,
	\beq
	\tilde{\psi}(p) = -\lambda \frac{\psi(0)}{ p^2 + \epsilon_B} \; .
	\label{Eq:tilde-psi-p}
	\eeq
	Fourier transforming back to coordinate space,
	one obtains the full wave function,
	\beq
	\psi(x) = \psi(0) e^{-|x| \sqrt{ \epsilon_B}} \; .
	\label{Eq:psi-x}
	\eeq
	In Eqs.~(\ref{Eq:tilde-psi-p}) and (\ref{Eq:psi-x}), the value of $\psi(0)$ is fixed by the condition~(\ref{Eq:psi-x-0}).
	
	\section{\label{app:scattering delta}Scattering states for the standard delta interaction}
	We review the scattering state sector of the standard delta-function potential in this appendix. 
	
	The momentum-space Schr\"{o}dinger equation~(\ref{Eq:Sch2Bdelta-momentum}) in the scattering sector involves using $E=k^2$, with $k$ real. The corresponding wave function can be derived by inversion of the  left-hand side operator of Eq.~(\ref{Eq:Sch2Bdelta-momentum}) in a manner similar to Eq.~(\ref{ansatz}), and takes the form
	\begin{equation}
	\tilde{\psi}(p) = 2\pi \delta(p-k) -\frac{\lambda}{p^2-k^2-i\epsilon}\psi(0)  \label{psipdelta}
	\; , 
	\end{equation}
	with similar physical interpretation (first term representing the incident wave and the $i\epsilon$ prescription for boundary conditions for outgoing scattered waves). 
	
	The reflection and transmission amplitudes $R$ and $T$ can be found by comparison against the standard asymptotic form of Eq.~(\ref{psixdeltaansatz}). Multiplying the momentum wave function of Eq. (\ref{psipdelta}) by $dp/2\pi$ and integrating both sides, we get
	\begin{equation} \label{Ypsizero}
	\psi(0)= \frac{1}{  1+ i\lambda/2k } \; .
	\end{equation}
	Now, using the result Eq.~(\ref{Ypsizero}) to perform an inverse Fourier transform on Eq. (\ref{psipdelta}),
	we can get the wave function in position space,
	\begin{align}
	\int \frac{dp}{2\pi} \tilde{\psi}(p) e^{ipx} &= e^{ikx} - \lim\limits_{\epsilon\rightarrow 0} \, \lambda \psi(0)\int  \frac{dp}{2\pi} \frac{e^{ipx}}{p^2-k^2-i\epsilon}\\ \nonumber
	\psi(x) &= e^{ikx} + \frac{\lambda}{2ik-\lambda} e^{ik|x|} \label{compactpsixdelta}\\ \nonumber
	&=e^{ikx}-\frac{\kappa}{\kappa + ik}e^{ik|x|} \; ,
	\end{align}
	where $\kappa =\sqrt{\epsilon_B}= -\lambda/2$ is the bound-state wave vector for the delta potential, and as expected the scattered wave is an outgoing 1D spherical wave. We can rewrite the wave function as
	\begin{align}
	\psi(x)&=e^{ikx}-\frac{\kappa}{\kappa + ik}e^{ik|x|}\\ \nonumber
	&=\begin{cases} 
	\displaystyle
	e^{ikx}-\frac{\kappa}{\kappa + ik}e^{-ikx}&x<0 \\ 
	\\
	\displaystyle
	\left(1-\frac{\kappa}{\kappa + ik}\right)e^{ikx}& x\geq 0 \; .\end{cases}
	\end{align}
	Hence, comparing the wave function~(\ref{compactpsixdelta}) with the form of Eq.~(\ref{psixdeltaansatz}) in the different regimes we get
	\begin{align}
	R &= -\frac{\kappa}{\kappa + ik}\\ \nonumber
	T &=  \frac{ik}{\kappa + ik} \; .
	\end{align} 
	
	We can also extract the phase shifts by diagonalizing the S-matrix that, for a symmetric potential, takes the form
	\begin{equation}
	S=\begin{pmatrix} T & R\\ R & T \end{pmatrix}\rightarrow \begin{pmatrix} e^{2i\delta_s} & 0\\ 
	0 & e^{2i\delta_p} \end{pmatrix}=\begin{pmatrix} 
	\f{-\kappa+i k}{\kappa+ik}& 0 \\ 0 & 1 \end{pmatrix},\end{equation}
	which leads respectively to the s- and p-phase shifts
	\begin{align} \label{eqnPhaseShift}
	\delta_s \equiv \delta_{0} &=\arctan \left(\frac{\kappa}{k}\right) \\ 
	\delta_p \equiv \delta_{1} &=0 \; .
	\end{align}

	\smallskip
	
	\section{\label{app:1DDeltaQFT}NRQFT treatment for the standard delta interaction}
	In this appendix, we use the method of Sec.~\ref{sec:NRQFTDeltaPrime} for a delta-function interaction. The vertex contribution for this interaction is given by
	\begin{equation}
	-i T_{\rm (tree)}^{(\delta)} = -i\lambda. 
	\end{equation}
	The first-order correction to this will include a non-zero contribution from the 1-loop diagram (similar to Fig.~\ref{fig:loop diagrams}b) given by
	\begin{align}
	-i T_{(1)}^{(\delta)} (Q) &= \int \frac{dkd\omega}{(2\pi)^2} \; \frac{i(-i\lambda)}{\omega-k^2/2+i\epsilon} \frac{i(-i\lambda)}{(Q-\omega)-k^2/2+i\epsilon}  \label{eq:1-loop integral delta} \\
	& = - \frac{\lambda^2}{2\sqrt{Q}} \label{eq:T1Delta}
	\; ,
	\end{align}
	where $Q$ is the COM energy of the system. We can now find all the higher-order loop contributions by following this same procedure recursively. Summing up the $T$-matrix values up to all orders, we obtain
	\begin{equation}
	T_{\rm (exact)}^{(\delta)} = \frac{\lambda }{1+\frac{i\lambda/2}{\sqrt{Q}}} \label{eq:TmatrixDelta_1}
	\; .
	\end{equation}
	For negative $Q$, this T-matrix has a pole at the (negative) bound-state energy of magnitude $\epsilon_B = \lambda^2/4$ which agrees with the result obtained in \ref{sec:bound state delta}. 
	
	Using the expression for the bound-state energy, Eq.~(\ref{eq:TmatrixDelta_1}) can be rewritten as
	\begin{equation}
	T_{\rm (exact)}^{(\delta)} = \frac{-2\sqrt{\epsilon_B}}{1-\frac{i\sqrt{\epsilon_B}}{\sqrt{Q}}}
	\label{eq:TmatrixDelta_2}
	\; ,
	\end{equation}
	which has the same form and hence the same residue and pole as the derivative-delta interaction obtained in Sec.~\ref{sec:NRQFTDeltaPrime}.
	
	In short, as the delta-function potential is not scale invariant, there exists a characteristic scale that supports a ground state, as shown by Eq.~(\ref{eq:TmatrixDelta_1}); and this scale naturally governs the scattering behavior via Eq.~(\ref{eq:TmatrixDelta_2}). Therefore, there is no dimensional transmutation and, thus, no need for a running coupling and for renormalization. As a consequence of these properties, we conclude that the delta-function model is distinctly different from the derivative-delta potential discussed in the main body of this paper.
	
	\section{Universal relations}
	\label{sec:Universal relations}
	
	As anticipated in Subsec.~\ref{sec:Tan-contact}, the contact determines the short-distance behavior of theories with short-range interactions. Thus, the analogue of the well-known set of universal relations can be expected to exist for the present system. Following the derivations of Ref.~\cite{WernerCastin}, we expect the behavior of the $N$-particle wave function to satisfy the asymptotic relation
	\beq
	\label{Eq:PsiAsymptotics}
	\begin{aligned}
		\!\!\!\!\
		\psi(x_1,x_2,\dots,x_N)  
		\underset{(r_{ij} \rightarrow 0 ) }{\sim} &
		f(r_{ij} \sqrt{\epsilon_B}) \\ 
		& \!\!\!\!\!  
		\times A_{ij}(R_{ij},\{x_q\}) + O(r_{ij}) 
		\; ,
	\end{aligned}
	\eeq
	where $f(x)$ is the governing function for the short-distance asymptotic behavior ($r_{ij}\sim 0 $) of the two-body wave function with respect to the distance $r_{ij}$ between particles $i,j$, with each index corresponding to a different particle species. In Eq.~(\ref{Eq:PsiAsymptotics}), $\{x_q\} =(x_q)_{q\neq i,j}$; $R_{ij} = (x_i + x_j)/2$ is the center-of-mass coordinate of particles $i,j$; and $A_{ij}$ is the regular part of the wave function. The above form is expected at short distances (as a result of the short-range behavior of the two-body problem), i.e. for $r_{ij} \sqrt{\epsilon_B} \ll 1$.
	
	A more complete derivation of universal relations will be given elsewhere. However, here we outline how
	one would proceed for the dimer or pair distribution function (density-density correlation function), which is given by
	\bea
	\! \! \! \! \! \!  \!\!  
	g_{12}^{(2)}(R, r) &&= \int dx_1 \dots dx_{N} \, \bigl| \psi(x_1,x_2,\dots,x_N) \bigr|^2 \\
	&& \!\!\!\!\!\! \times \sum_{i_1,i_2}^{N_1,N_2} \delta(x_{i_1} - f_1)\delta(x_{i_2} - f_2) \; , \nonumber
	\eea
	where $i_1, i_2$ vary over the particles in each of the two different species, $f_1 = R + r/2$, and $f_2 = R - r/2$. 
	Here, for a given pair of particles $1,2$, we are suppressing the indices of the center-of-mass ($R_{12}$) and relative ($r_{12}$) distances. Then, inserting Eq.~(\ref{Eq:PsiAsymptotics}) into the above expression for $g_{12}^{(2)}(R, r)$, we find that,	when $r \to 0$, each term in $g_{12}^{(2)}(R,r)$ is seen to be dominated by the divergent part of the wave function. 
	Thus,
	\bea
	g_{12}^{(2)}(R, r) \underset{(r \rightarrow 0 ) }{\sim} 
	f^2(r \sqrt{\epsilon_B}) \; F(R) \; ,
	\eea
	where
	\bea
	F(R) = \!\!\!\sum_{i_1,i_2}^{N_1,N_2}\!\!\! \int \!\!\!\prod_{k\neq i_1,i_2}\!\!\!\!\! dx_k  | A_{i_1,i_2} (R,\{x_q\}) |^2 \; ,
	\eea
	with $\{x_q\} =(x_q)_{q\neq i,j}$. Finally, integrating over $R$, the spatially integrated dimer distribution function
	$G_{12}^{(2)}(r)$ is obtained, which takes its short-distance behavior from $g_{12}^{(2)}$, i.e.
	\beq
	G_{12}^{(2)}(r) \equiv \int dR \; g_{12}^{(2)}(R,r) 
	\underset{(r \rightarrow 0 ) }{\sim} 
	f^2 (r\sqrt{\epsilon_B}) \; \mathcal J \; .
	\eeq
	The constant $\mathcal J$ is determined by the contact (up to a coupling-independent factor).
	The contact can also be derived from Eq.~(\ref{Eq:PsiAsymptotics}) 
	by evaluating the expectation value of the Hamiltonian. The central point of the above derivation is that $\sqrt{\epsilon_B}$ sets the scale 
	for the short-distance behavior.  Therefore, once the divergent piece has been factored out, the remaining factors are proportional to 
	$| A_{i_1,i_2} (R,\{x_q\}) |^2$.

\end{appendix}


\end{document}